\documentclass[Physsubmission, Phys]{SciPost}

\hypersetup{
    colorlinks,
    linkcolor={red!50!black},
    citecolor={blue!50!black},
    urlcolor={blue!80!black}
}

\usepackage{lineno}
\usepackage{enumitem}
\usepackage[bitstream-charter]{mathdesign}
\DeclareSymbolFont{usualmathcal}{OMS}{cmsy}{m}{n}
\DeclareSymbolFontAlphabet{\mathcal}{usualmathcal}

\newcommand{\pp}{\ensuremath{\mathrm {p\kern-0.05em p}}}
\newcommand{\PbPb}{\ensuremath{\mbox{Pb--Pb}}}
\newcommand{\GeVc}{\ensuremath{\mathrm{GeV}\kern-0.05em/\kern-0.02em c}}

\newcommand{\pT}{\ensuremath{p_{\mathrm{T}}}}
\newcommand{\kT}{\ensuremath{k_{\mathrm{T}}}}
\newcommand{\pTjet}{\ensuremath{p_{\mathrm{T,\;jet}}}}

\newcommand{\pTfullsubjet}{\ensuremath{p_{\mathrm{T,\;subjet}}}}

\newcommand{\zr}{\ensuremath{z_r}}

\begin{document}

\setlength{\abovedisplayskip}{6pt}
\setlength{\belowdisplayskip}{6pt}

\begin{center}{\Large \textbf{Measurements of sub-jet fragmentation with ALICE}}\end{center}

\begin{center}
James Mulligan\textsuperscript{1,2$\star$} for the ALICE Collaboration
\end{center}

\begin{center}
{\bf 1} Nuclear Science Division, Lawrence Berkeley National Laboratory, Berkeley, California, USA
\\
{\bf 2} Physics Department, University of California, Berkeley, CA, USA
\\
* james.mulligan@berkeley.edu
\end{center}

\begin{center}
\today
\end{center}


\definecolor{palegray}{gray}{0.95}
\begin{center}
\colorbox{palegray}{
  \begin{tabular}{rr}
  \begin{minipage}{0.1\textwidth}
    \includegraphics[width=30mm]{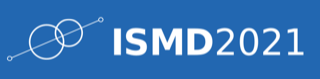}
  \end{minipage}
  &
  \begin{minipage}{0.75\textwidth}
    \begin{center}
    {\it 50th International Symposium on Multiparticle Dynamics}\\ {\it (ISMD2021)}\\
    {\it 12-16 July 2021} \\
    \doi{10.21468/SciPostPhysProc.?}\\
    \end{center}
  \end{minipage}
\end{tabular}
}
\end{center}

\section*{Abstract}
{\bf

High-energy jets offer rich opportunities to study quantum chromodynamics,
from investigating the limits of perturbative calculability
to constraining the emergent properties of the quark-gluon plasma (QGP).
In these proceedings, we present new measurements of the fragmentation properties of jets. 
We report distributions of the sub-jet momentum fraction \zr{} measured in 
pp and Pb--Pb collisions with ALICE at the Large Hadron Collider. 
These measurements serve as input to test the universality of jet fragmentation in the QGP,
and offer a path to elucidate jet quenching effects in the large-$z$ region.
}

\section{Introduction}

Jet measurements offer opportunities to test perturbative calculations
in quantum chromodynamics and to probe the properties of the 
QGP \cite{Larkoski_2020, TheBigPicture}.
In these proceedings, we consider measurements of \textit{sub-jets},
defined by first inclusively clustering jets with the 
anti-\kT{} algorithm \cite{antikt} with radius $R$,
and then reclustering the jet constituents with the anti-\kT{} algorithm with sub-jet radius $r<R$ \cite{Dai:2016hzf}.
We focus on the fraction of transverse momentum carried by the sub-jet:
\[
\zr = \frac{\pTfullsubjet}{\pTjet}.
\]
In \pp{} collisions, both the inclusive and leading sub-jet \zr{} distributions have been calculated 
perturbatively \cite{Kang:2017mda, Neill:2021std}.
These calculations suggest several interesting features that can be tested by experimental data:
the role of threshold resummation in the large-\zr{} region
and, in the leading sub-jet case,
nonlinear evolution of the jet fragmentation function in the perturbative calculation.
In heavy-ion collisions, sub-jets have been proposed as
sensitive probes of jet quenching \cite{Kang:2017mda, Neill:2021std, Apolinario:2017qay, Caucal:2020xad}.
The sub-jet \zr{} observable presents several unique opportunities:
\begin{enumerate}[itemsep=-0.5ex,topsep=4pt]
    \item \textit{Test the universality of jet fragmentation in the QGP}. 
    In vacuum, it is expected that the parton-to-jet fragmentation function, $J(z)$,
    is equal to the parton-to-subjet fragmentation function $J_{r}(z)$.
    However, it is unknown whether the universality of jet fragmentation functions holds in the QGP \cite{Qiu:2019sfj}.
    Measurements of \zr{} are directly sensitive to $J_{r,\rm{med}}(z)$, and can be used to extract it. 
    The extracted $J_{r,\rm{med}}(z)$ can then be compared
    to an independently extracted $J_{\rm{med}}(z)$
    to test the universality of in-medium jet fragmentation.
    
    \item \textit{Probe high-$z$ fragmentation}.
    Sub-jet fragmentation is complementary to the longitudinal
    momentum fraction of hadrons in jets \cite{ATLAS:2019dsv, CMS:2014jjt}.
    Sub-jet measurements offer the benefit of probing higher $z$ than hadron measurements,
    and, in doing so, offer the possibility to access a quark-dominated sample of jets and expose the interplay of soft medium-induced radiation with the relative suppression of gluon vs. quark jets.

    \item \textit{Measure sub-jet energy loss at the cross-section level}.
    Recently, a well-defined method of measuring out-of-cone energy loss at the cross-section level was proposed, by computing moments of the leading sub-jet \zr{} distribution \cite{Neill:2021std}.
    This ``sub-jet energy loss'', describing the fraction of jet \pT{} not carried by the leading sub-jet, can then be computed in both \pp{} and \PbPb{} collisions,
    and contrasted with other measures of jet modification.
    
\end{enumerate}

\section{Results}

All presented results use $R=0.4$ jets reconstructed from charged particles with pseudorapidity $|\eta|<0.9$,
and are corrected for detector effects and (in \PbPb{} collisions) underlying-event fluctuations.

\begin{figure}[!b]
\centering{}
\includegraphics[scale=0.34]{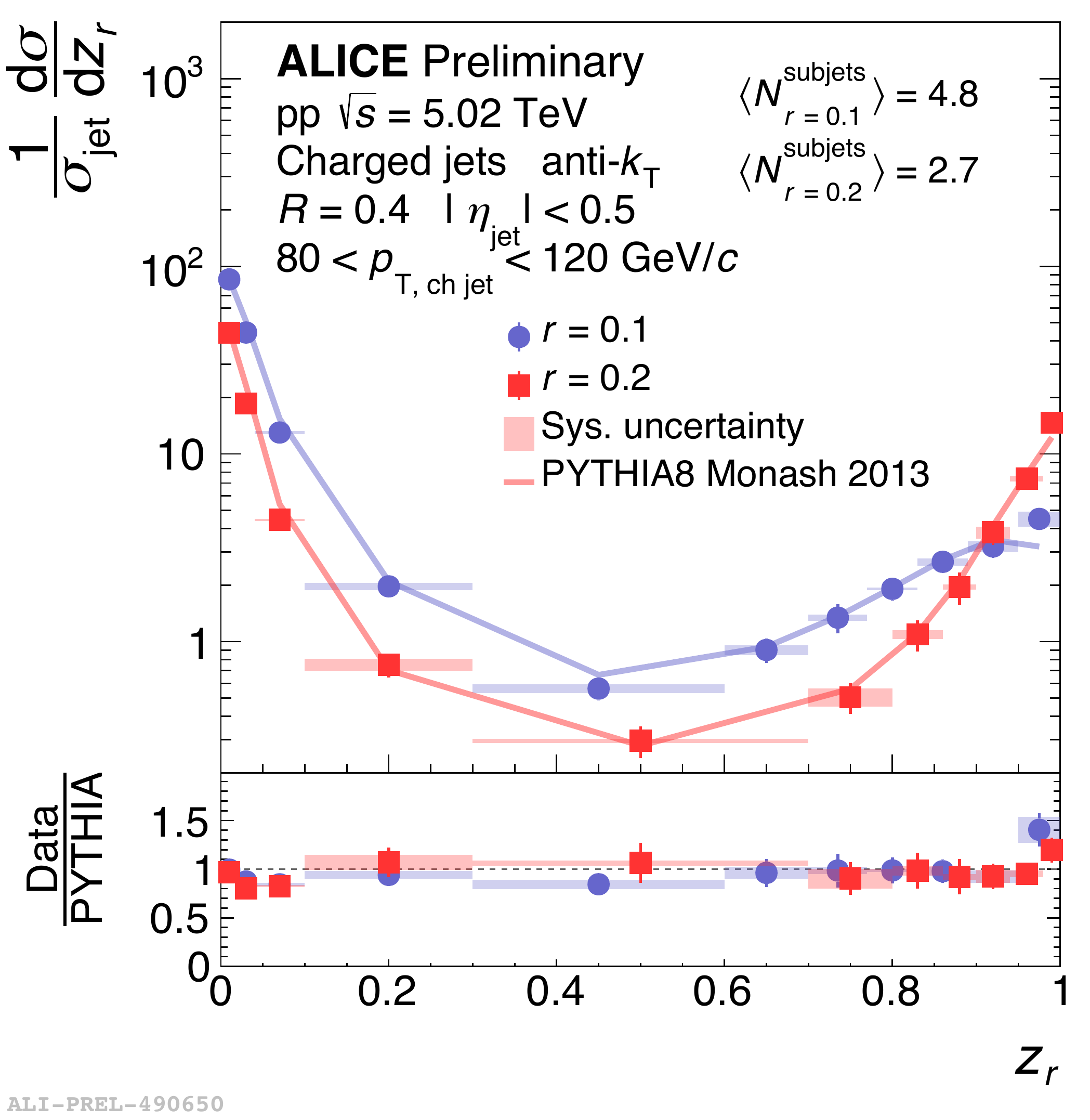}
\includegraphics[scale=0.34]{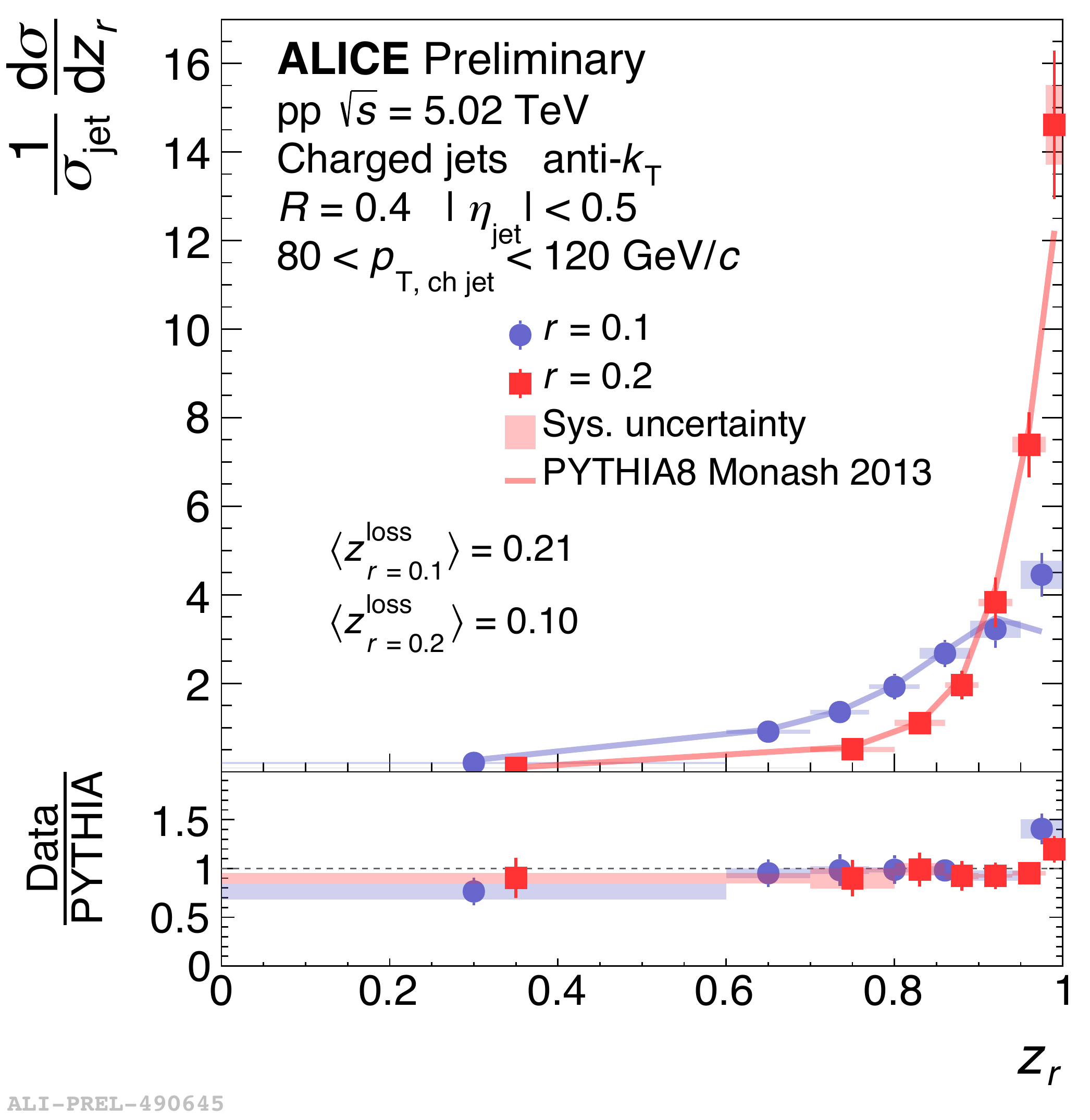}
\caption{ALICE measurements of inclusive (left) and leading (right) sub-jet \zr{} distribution 
in \pp{} collisions for two different sub-jet radii, compared to PYTHIA8 \cite{pythia}.}
\label{fig:zr-pp}
\end{figure}

\subsection{Sub-jet fragmentation in proton-proton collisions}

Figure \ref{fig:zr-pp} shows the measured \zr{} distributions for inclusive (left) 
and leading (right) sub-jets. 
The \zr{}-differential cross sections are normalized
such that their integrals are equal to the average number of sub-jets per jet.
For $\zr>0.5$ the leading and inclusive distributions are identical.
As \zr{} becomes small, the inclusive sub-jet distribution grows due to soft radiations emitted
from the leading sub-jet, whereas the leading sub-jet distribution falls to zero.
The distributions are generally described well by PYTHIA8 Monash 2013~\cite{pythia,Skands:2014pea}, however
there is disagreement at large \zr{} – this may be due to
threshold resummation (which is not directly included in PYTHIA8) or to hadronization effects.
Using the leading sub-jet distributions, we also compute the ``sub-jet energy loss'':
\[
\langle z_{\rm{loss}}\rangle =1-\int _0^1dz_r\;z_r \frac{1}{\sigma}\frac{d\sigma}{dz_r},
\]
which describes the fraction of \pT{} inside the jet that is not contained within
the leading subjet \cite{Neill:2021std}. 
We find that $\langle z_{\rm{loss}}\rangle = 0.21$ for $r=0.1$
and decreases to $\langle z_{\rm{loss}}\rangle = 0.10$ for $r=0.2$.

\subsection{Sub-jet fragmentation in Pb--Pb collisions}

\begin{figure}[!b]
\centering{}
\includegraphics[scale=0.34]{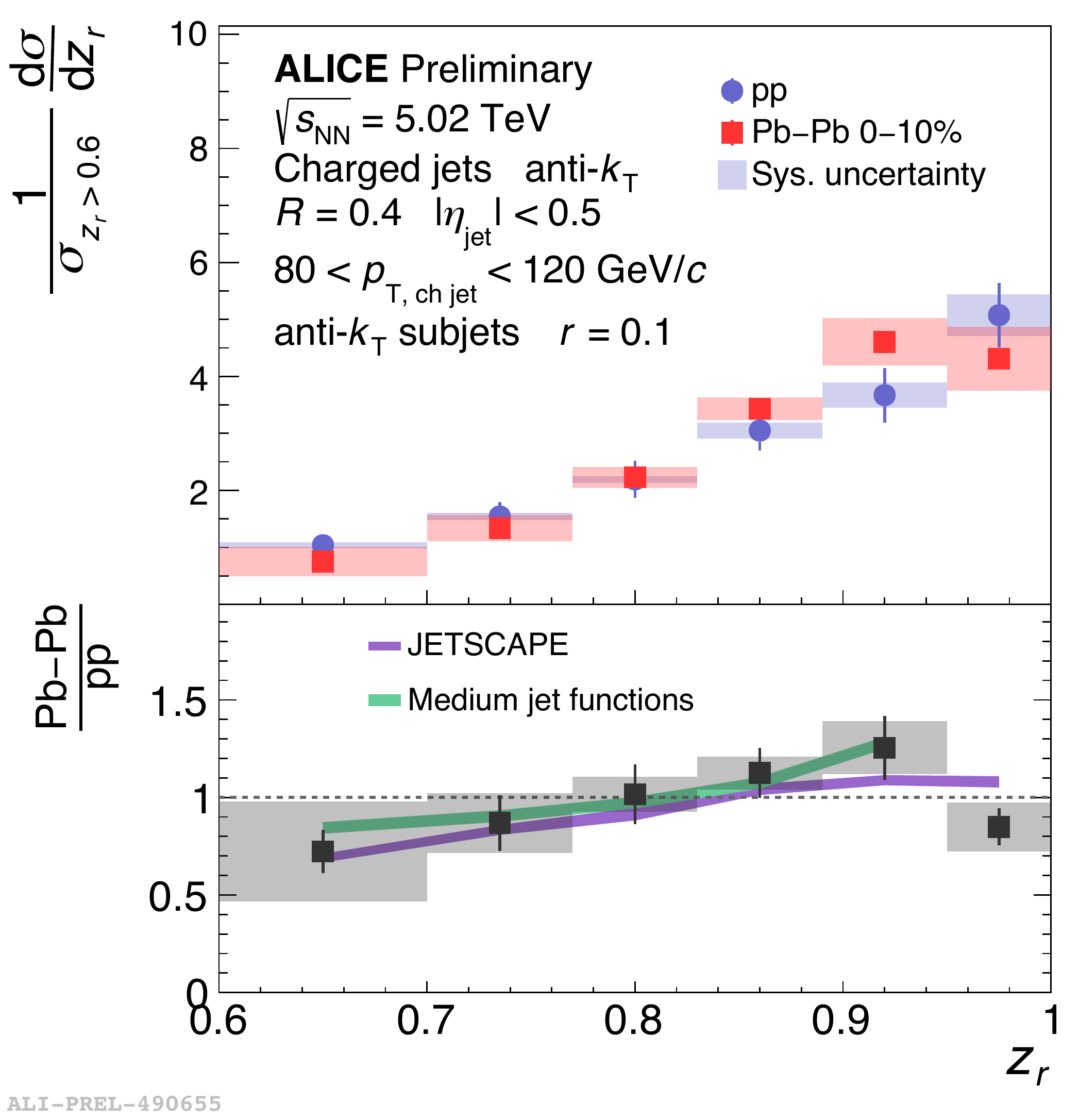}
\includegraphics[scale=0.34]{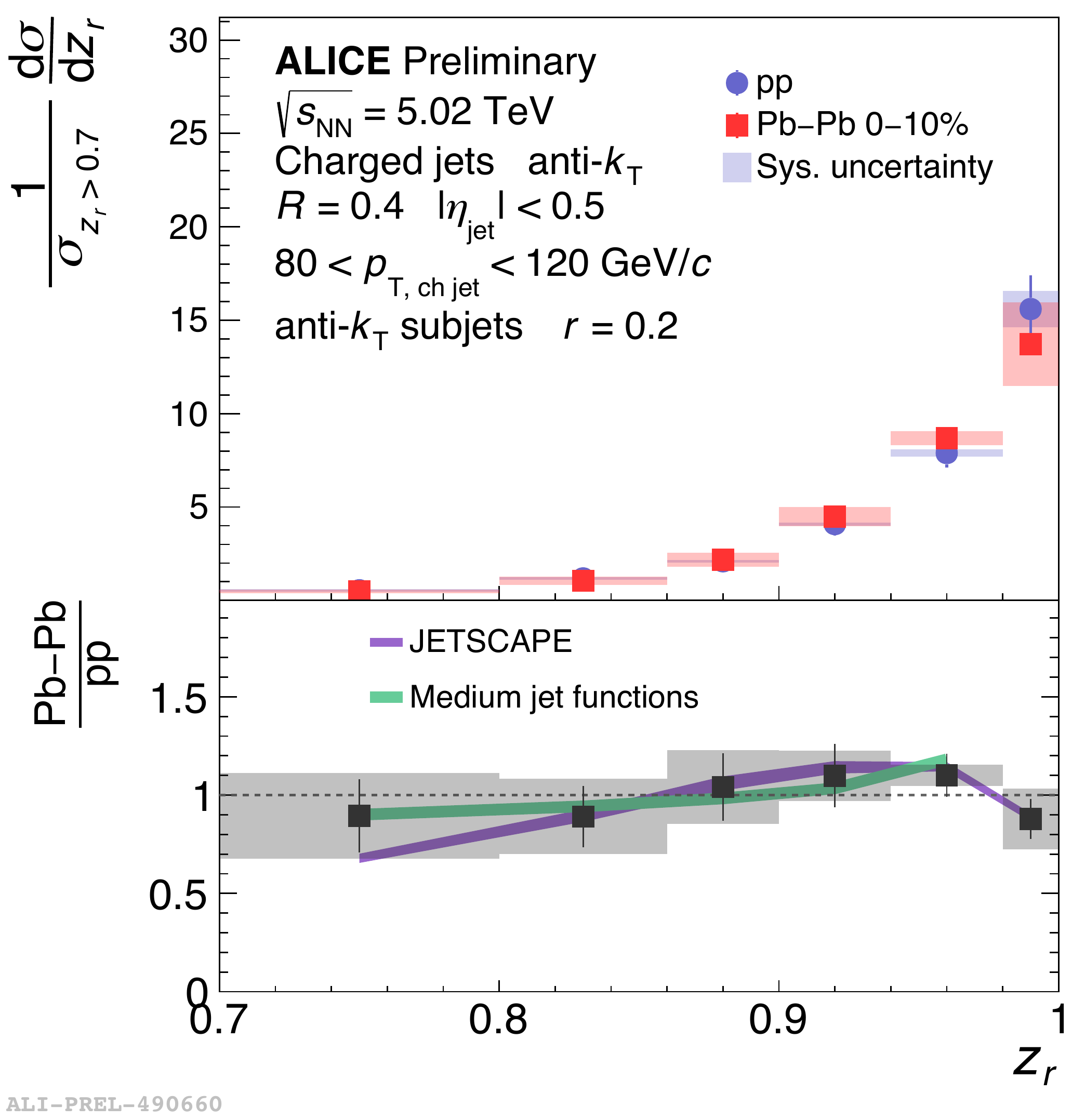}
\caption{Measurements of sub-jet fragmentation for sub-jet radii $r=0.1$ (left) 
and $r=0.2$ (right) in \pp{} and \PbPb{} collisions, 
compared to predictions \cite{Kang:2017mda, Qiu:2019sfj, Putschke:2019yrg, LBT, Majumder_2013}.}
\label{fig:zr-AA}
\end{figure}

The fluctuating underlying event in heavy-ion collisions poses
an additional challenge, since it can alter the number of reconstructed sub-jets.
To simplify this problem, we focus on leading sub-jets at large \zr{}.\footnote{Even with this restriction, underlying event fluctuations can
cause the leading sub-jet to be misidentified,
in analogy to groomed jet observables \cite{Mulligan:2020tim},
although with improved robustness to mistagging effects \cite{STAR:2021kjt}.}
Figure \ref{fig:zr-AA} shows the \zr{} distributions 
in \pp{} and \PbPb{} collisions for $r=0.1$ (left) and $r=0.2$ (right). 
For $r=0.1$, the distributions are consistent with a mild hardening effect
in \PbPb{} compared to \pp{} collisions, which reverses as $\zr \rightarrow 1$. 
These results are compared to JETSCAPE 
\cite{Putschke:2019yrg, LBT, Majumder_2013}
and SCET-based calculations \cite{Kang:2017mda, Qiu:2019sfj},
both of which generally describe the data well.
To understand the behavior of the data, note that in vacuum there are significant
differences in the parton-to-subjet fragmentation functions 
between quarks and gluons \cite{Neill:2021std}.
If the QGP suppresses gluon jets more than quark jets,
a hardening effect of the \zr{} distribution would be expected
– in line with previous measurements of hadron fragmentation \cite{Spousta:2015fca}.
On the other hand, medium-induced soft radiations
can shift the distribution to smaller \zr{}.
This competition can give non-trivial modification patterns.
In particular, as $\zr \rightarrow 1$, the jet sample in vacuum
becomes almost entirely dominated by quark jets – exposing a region depleted by 
soft medium-induced emissions, which is consistent with our observations.
 
\section{Conclusion}

We have presented new measurements of sub-jet fragmentation with ALICE. 
In proton-proton collisions, these measurements provide opportunities to test
non-linear evolution of jet fragmentation functions and the role of threshold resummation.
In heavy-ion collisions, these measurements serve as a key ingredient to test the universality
of jet fragmentation in the QGP. 
By probing large \zr{}, these measurements isolate a region of quark-dominated jets that
may expose a region depleted by medium-induced soft radiation.
Future measurements of \zr{} in coincidence with other substructure observables
such as the groomed jet radius \cite{ALICE:2021obz} offer the potential to disentangle this effect
from the relative suppression of gluon jets to quark jets.


\bibliography{main.bib}
\nolinenumbers

\end{document}